\documentclass[seceq,supplement]{ptptex}

\def\lsim{\mathrel{\rlap{\lower4pt\hbox{\hskip1pt$\sim$}}
    \raise1pt\hbox{$<$}}}                
\def\gsim{\mathrel{\rlap{\lower4pt\hbox{\hskip1pt$\sim$}}
    \raise1pt\hbox{$>$}}}                
    
\usepackage{amsmath}
\usepackage{amssymb}
\usepackage{amsfonts}
\usepackage{subfigure}
\usepackage{graphicx}

\def\square{\vcenter{\vbox{\hrule height.4pt
          \hbox{\vrule width.4pt height8pt
          \kern8pt\vrule width.4pt}\hrule height.4pt}}}

\newcommand{\beq}{\begin{equation}}
\newcommand{\eeq}{\end{equation}}
\newcommand{\bqa}{\begin{eqnarray}}
\newcommand{\eqa}{\end{eqnarray}}

\title{Hard-thermal-loop QCD Thermodynamics}

\author{Michael Strickland$^{1,2}$, Jens O. Andersen$^3$, Lars E. Leganger$^3$, and Nan Su$^2$}
\inst{
$^1$ Department of Physics, Gettysburg College, Gettysburg, PA 17325, 
USA \\
$^2$ Frankfurt Institute for Advanced Studies, Ruth-Moufang-Str. 1, 
D-60438 Frankfurt am Main, Germany \\
$^3$ Department of Physics, Norwegian University of Science
and Technology, H{\o}gskoleringen 5, N-7491 Trondheim, Norway
}

\abst{
Naively resummed perturbative approximations to the thermodynamic functions
of QCD do not converge at phenomenologically relevant temperatures.  Here
we review recent results of a three-loop hard-thermal-loop perturbation
theory calculation of the thermodynamic functions of a quark-gluon plasma for general
$N_c$ and $N_f$.  We show comparisons of our recent results with lattice data from
both the hotQCD and Wuppertal-Budapest groups.  We demonstrate that the three-loop
hard-thermal-loop perturbation result for QCD thermodynamics agrees with
lattice data down to temperatures $T\sim2\,T_c$.
}

\begin{document}

\maketitle

\section{Introduction}

A key question in the study of the quark-gluon plasma (QGP) is whether or not
one can use a weakly-coupled framework to calculate properties of the
system at temperatures near the QGP phase transition temperature, $T_c\sim170$ MeV.
This is phenomenologically relevant because the ultrarelativistic heavy-ion collision
experiments at Brookhaven National Labs (RHIC), and forthcoming at CERN (LHC) 
generate initial QGP temperatures which are on the order of $2\,T_c$ and $5\,T_c$,
respectively.  At these temperatures the strong coupling constant is approximately
$\alpha_s=g^2_s/4\pi \sim 0.3$.  Unfortunately, this places the system in regime
where the coupling is neither very small nor very large.  Prior to experiments at
RHIC, theorists expected that the QGP could be described in terms of weakly interacting
quasiparticles; however, data from RHIC  suggested that the state of matter created
behaved more like a strongly coupled fluid with a small 
viscosity~\cite{rhicexperiment}. This has inspired work on strongly-coupled
formalisms based on e.g. the AdS/CFT correspondence.
However, some observables such as jet quenching~\cite{pert}
and elliptic flow~\cite{Xu:2007jv} can also be described using perturbative methods and 
so it is difficult to decide whether the plasma is strongly or weakly coupled based only on 
RHIC data.

In the upcoming heavy-ion collisions at LHC, the energy densities and therefore
the initial temperatures will be higher than those achieved at 
RHIC. One expects initial temperatures
on the order of  $5\,T_c$. An important question is then whether the matter generated
can be described in terms of weakly interacting quasiparticles
at these higher temperatures.
Lattice simulations of QCD provide a clean testing ground for 
the quasiparticle picture and in this proceedings report we compare recently
obtained next-to-next-to-leading order (NNLO) results for thermodynamic functions
of QCD\cite{Andersen:2010wu}  with lattice data~\cite{hotqcd1,borsy}
and with previous results at leading order (LO) and 
next-to-leading order (NLO)~\cite{htlpt}.
The calculations presented here are based on the hard-thermal-loop 
perturbation theory (HTLpt) reorganization of finite-temperature
perturbation theory. In HTLpt one expands around an ideal gas of massive
gluonic and quark
quasiparticles where screening effects and Landau damping are gauge-invariantly built in.
Our results indicate that lattice data are consistent with 
a quasiparticle picture down to temperatures of $T\sim2\,T_c$.

The free energy of QCD is now
known up to order $\alpha_s^3\log\alpha_s$~\cite{npert,BN}.
Unfortunately, a straightforward application of perturbation theory 
is of no quantitative use at phenomenologically relevant temperatures.
The problem is that the weak-coupling expansion oscillates wildly and shows
no sign of convergence unless the temperature is extremely high.
For example, if one compares the $g_s^3$-contribution to the QCD free energy
with three quark flavors to the $g^2_s$-contribution, the former is smaller
only if $\alpha_s\leq0.07$, which corresponds to $T\sim10^5$
GeV or $T\sim5\times10^5\,T_c$.
There are several ways of reorganizing the perturbative series at finite
temperature~\cite{reviews}.  These reorganizations are based on a
quasiparticle picture where one is perturbing about an ideal gas of 
massive
quasiparticles, rather than that of an ideal gas of massless quarks and gluons.
In scalar $\phi^4$-theory the basic idea
is to add and subtract a thermal mass term
from the bare Lagrangian and to include the added piece in the free part of 
the Lagrangian. The subtracted piece is then treated as an interaction
on the same footing as the quartic term~\cite{spt}.
In gauge theories, however, simply adding and subtracting a local mass
term, violates gauge 
invariance~\cite{gaugebreak}. 
Instead, one adds and subtracts
an HTL improvement term, which dresses the propagators and vertices
self-consistently so that the reorganization is manifestly gauge 
invariant~\cite{htlaction}.

\section{Hard-thermal-loop perturbation theory}
The Lagrangian density for an ${\rm SU}(N_c)$
Yang-Mills theory with $N_f$ fermions
in Minkowski space is
\bqa
{\cal L}_{\rm QCD}&=&
-{1\over2}{\rm Tr}\left[G_{\mu\nu}G^{\mu\nu}\right]
+i \bar\psi \gamma^\mu D_\mu \psi 
+{\cal L}_{\rm gf}
+{\cal L}_{\rm gh}
+\Delta{\cal L}_{\rm QCD}\;,
\label{L-QED}
\eqa
%
where the field strength is 
$G^{\mu\nu}=\partial^{\mu}A^{\nu}-\partial^{\nu}A^{\mu}-ig_s[A^{\mu},A^{\nu}]$
and the covariant derivative is $D^{\mu}=\partial^{\mu}-ig_sA^{\mu}$.
$\Delta{\cal L}_{\rm QCD}$ contains the counterterms necessary to cancel
the ultraviolet divergences.
The ghost term ${\cal L}_{\rm gh}$ depends on the gauge-fixing term
${\cal L}_{\rm gf}$. In this paper we choose the class of covariant gauges
where the gauge-fixing term is ${\cal L}_{\rm gf}=-{1\over\xi}{\rm Tr}
\left[\left(\partial_{\mu}A^{\mu}\right)^2\right]$.  With the standard normalization, we
have $c_A=N_c$, $d_A=N_c^2-1$, $s_F=N_f/2$, $d_F=N_cN_f$,
and $s_{2F}=(N_c^2-1)N_f/4N_c$.

Hard-thermal-loop perturbation theory is a reorganization
of the perturbation
series for thermal QCD. The Lagrangian density is written as
\bqa
{\cal L}= \left({\cal L}_{\rm QCD}
+ {\cal L}_{\rm HTL} \right) \Big|_{g_s \to \sqrt{\delta} g_s}
+ \Delta{\cal L}_{\rm HTL}\;,
\label{L-HTLQCD}
\eqa
where $\Delta{\cal L}_{\rm HTL}$ contains the additional counterterms 
necessary to cancel the ultraviolet divergences introduced by HTLpt.
The HTL improvement term is
\bqa
\hspace{-3mm}
{\cal L}_{\rm HTL}=-{1\over2}(1\!-\!\delta)m_D^2 {\rm Tr}\!
\left(\!G_{\mu\alpha}\!\left\langle\!{y^{\alpha}y^{\beta}\over(y\cdot D)^2}
	\!\!\right\rangle_{\!\!y}\!G^{\mu}_{\;\;\beta}\!\right)
        \!+(1\!-\!\delta)\,i m_q^2 \bar{\psi}\gamma^\mu 
\left\langle\! {y^{\mu}\over y\cdot D}
	\!\right\rangle_{\!\!y}\psi , 
\label{L-HTL}
\eqa
where $y^{\mu}=(1,\hat{{\bf y}})$ is a light-like four-vector,
and $\langle\ldots\rangle_{ y}$
represents the average over the directions
of $\hat{{\bf y}}$. The free parameters $m_D$ and $m_q$ are identified
with the Debye screening mass and the fermion thermal mass, respectively.
The parameter $\delta$ is a formal expansion parameter and bookkeeping device:
HTLpt is defined as an expansion in powers of $\delta$ around $\delta=0$.
This expansion generates systematically
dressed propagators and vertices. It also
automatically generates new higher-order terms that ensure that there is
no overcounting of Feynman diagrams.  We emphasize that 
HTLpt is gauge invariant at each order in $\delta$.

The HTL perturbative expansion generates
new divergences. We use ${\overline{\rm MS}}$ dimensional regularization 
with a renormalization scale $\mu$ to regularize
infrared and ultraviolet divergences. 
There is no rigorous proof that HTLpt is
renormalizable, so the general structure of the counterterms is unknown.
However, one can show that through NNLO, HTLpt can be renormalized using only
local counterterms for the vacuum, the Debye and fermion masses, and 
the coupling constant. The counterterm for $\alpha_s$ coincides with the
perturbative value giving rise to the standard one-loop running.
We do not list the counterterms here, but present the full
results elsewhere~\cite{finalhtl}.
If the expansion in $\delta$ could be carried out to all orders, the
final result would be independent of the HTL parameters $m_D$ and $m_q$.
However, at any finite order in $\delta$, the results depend on $m_D$ and $m_q$.
A prescription is then required to determine these parameters.  We will
discuss the prescription we use below. 

\section{Thermodynamic potential}
In this section, we present the final results for the thermodynamic potential
$\Omega$ at orders $\delta^0$ (LO), $\delta$ (NLO), and $\delta^2$ (NNLO).
The LO and NLO results were first obtained previously~\cite{htlpt}
and they are listed here for completeness. At LO, the thermodynamic
potential was calculated exactly, while at NLO and NNLO the resulting
expressions for the diagrams are too complicated. To make the calculations
tractable, the thermodynamic
potential is therefore evaluated approximately by expanding them
in powers of $m_D/T$ and $m_q/T$ which assumes that these ratios are 
${\cal O}(g_s)$. This implies that the thermodynamic potential is
evaluated in a double expansion in $g_s$, $m_D/T$, and $m_q/T$, and we have 
kept terms that contribute naively through order $g_s^5$.
Due to the magnetic mass 
problem~\cite{lindeir},
HTLpt suffers from the same infrared divergences as ordinary perturbation
theory and $g_s^5$ is the highest order computable using only perturbative
methods.

The complete expression for the leading order thermodynamic potential
is given by~\cite {htlpt}
\bqa\nonumber
\frac{\Omega_{\rm LO}}{{\cal F}_{\rm ideal}} &=& 
1+{7\over4}{d_F\over d_A}
- {15 \over 2} \hat m_D^2 
-30{d_F\over d_A}\hat{m}_q^2
+ 30 \hat m_D^3
+ {45 \over 4}
\left(\log {\hat \mu \over 2}
        - {7\over2} + \gamma_E + {\pi^2\over 3} \right)
        \hat m_D^4 \\&&
-60{d_F\over d_A}(\pi^2-6)\hat{m}_q^4\;, 
\label{Omegaren-1}
\eqa
where ${\cal F}_{\rm ideal}=-(N_c^2-1)\pi^2T^4/45$ is the free energy of
an ideal gas of noninteracting gluons
and $\gamma_E$ is the 
Euler-Mascheroni constant. Moreover, we have introduced the dimensionless
parameters $\hat{\mu}=\mu/2\pi T$, $\hat{m}_D=m_D/2\pi T$, 
and $\hat{m}_q=m_q/2\pi T$.

The NLO thermodynamic potential reads~\cite {htlpt}
\bqa
{\Omega_{\rm NLO}\over{\cal F}_{\rm ideal}
}&=&
	1 
+{7\over4}{d_F\over d_A}
- 15 \hat m_D^3 
	- {45\over4}\left(\log\hat{\mu\over2}-{7\over2}+\gamma_E+{\pi^2\over3}
\right)\hat m_D^4+60{d_F\over d_A}(\pi^2-6)\hat{m}_q^4
\nonumber \\ && \nonumber
+	{c_A\alpha_s\over3\pi}
\left[-{15\over4}+45\hat m_D
-{165\over4}\left(\log{\hat\mu \over 2}-{36\over11}\log\hat{m}_D
-2.001\right)\hat m_D^2
\right.\nonumber \\ && \nonumber\left.
+{495\over2}\left(\log{\hat\mu \over 2}+{5\over22}+\gamma_E\right)\hat m_D^3
\right] 
\\ && \nonumber
+{s_F\alpha_s\over\pi}\left[-{25\over8}+15\hat m_D
+5\left(\log{\hat{\mu}\over2}-2.33452\right)\hat m_D^2
\right. \\ && \nonumber \left.
-30\left(
\log{\hat\mu \over 2}-{1\over2}+\gamma_E+2\log2\right)\!\!\hat m_D^3
\right.
\\ &&\left.
	-45\left(\log{\hat\mu \over 2}+2.19581\right)\hat m_q^2
+180\hat{m}_D\hat{m}_q^2\right]\;.
\label{Omega-NLO}
\eqa

Finally, our new result for the NNLO thermodynamic potential for QCD is
\begin{eqnarray}\nonumber
{\Omega_{\rm NNLO}\over{\cal F}_{\rm ideal}}
&=&
1+{7\over4}{d_F\over d_A}
-{15\over4}\hat{m}_D^3
\\ \nonumber  &&
+{c_A\alpha_s\over3\pi}\left[
-{15\over4}
+{45\over2}\hat{m}_D
-{135\over2}\hat{m}^2_D
-{495\over4}\left(\log{\hat\mu \over 2}+{5\over22}+\gamma_E\right)\hat m_D^3 
\right]
\\ && \nonumber
+{s_F\alpha_s\over\pi}\left[-{25\over8}+{15\over2}\hat{m}_D
+15\left(
\log{\hat\mu \over 2}-{1\over2}+\gamma_E+2\log2\right)\!\!\hat m_D^3
-90\hat{m}^2_q\hat{m}_D\right]
\\&& \nonumber
+\left({c_A\alpha_s\over3\pi}\right)^2\left[{45\over4}{1\over\hat{m}_D}
-{165\over8}\left(\log{\hat{\mu}\over2}-{72\over11}\log{\hat{m}_D}
-{84\over55}
-{6\over11}\gamma_E
\right.\right.\\ &&\left.\left. \nonumber
-{74\over11}{\zeta^{\prime}(-1)\over\zeta(-1)}
+{19\over11}{\zeta^{\prime}(-3)\over\zeta(-3)}
\right)
+{1485\over4}
\left(
\log{\hat{\mu}\over2}-{79\over44}+\gamma_E+\log2-{\pi^2\over11}
\right)\hat{m}_D
\right]
\\&&\nonumber
+\left({c_A\alpha_s\over3\pi}\right)\left({s_F\alpha_s\over\pi}\right)
\left[
{15\over2}{1\over\hat{m}_D}
-{235\over16}\left(\log{\hat{\mu}\over2}
-{144\over47}\log{\hat{m}_D}
\right.\right.\\ &&\nonumber\left.\left.
-{24\over47}\gamma_E
+{319\over940}+{111\over235}\log2
-{74\over47}{\zeta^{\prime}(-1)\over\zeta(-1)}
+{1\over47}{\zeta^{\prime}(-3)\over\zeta(-3)}
\right)
\right. \\ &&\nonumber\left.
+{315\over4}\left(\log{\hat{\mu}\over2}-{8\over7}\log2+\gamma_E+{9\over14}
\right)\hat{m}_D
+90{\hat{m}_q^2\over\hat{m}_D}
\right]
+\left({s_F\alpha_s\over\pi}\right)^2
\left[{5\over4}{1\over\hat{m}_D}
\right. \\ &&\nonumber \left.
+{25\over12}\left(
\log{\hat{\mu}\over2}+{1\over20}+{3\over5}\gamma_E-{66\over25}\log2
+{4\over5}{\zeta^{\prime}(-1)\over\zeta(-1)}
-{2\over5}{\zeta^{\prime}(-3)\over\zeta(-3)}\right)
\right.\\ && \left. \nonumber
-15\left(\log{\hat{\mu}\over2}
-{1\over2}+\gamma_E+2\log2
\right)\hat{m}_D
+30{\hat{m}_q^2\over\hat{m}_D}
\right]
\\ &&
+s_{2F}\left({\alpha_s\over\pi}\right)^2\left[{15\over64}(35-32\log2)
-{45\over2}\hat{m}_D\right]\;,
\label{Omega-NNLO}
\end{eqnarray}
where $\zeta(z)$ is the Riemann zeta-function.

As pointed out earlier, the HTL mass parameters are completely arbitrary and
we need a prescription for them in order to complete a calculation.
The variational mass prescription unfortunately gives rise to 
a complex Debye mass and $m_q=0$ at NNLO. 
One strategy is therefore to throw away the 
imaginary part of the thermodynamic potential to obtain
thermodynamic functions that are 
real valued~\cite{qednnlo,pg}. 
Here we use another strategy explored in Refs.~\citen{qednnlo,pg}
that is inspired by dimensional reduction:
We equate the Debye mass with the mass parameter of 
three-dimensional electric QCD (EQCD)~\cite{BN}, i.e.~$m_D=m_E$. In 
Ref.~\citen{BN}, it was calculated to NLO giving
\bqa\nonumber
m_D^2&=&{4\pi\alpha_s\over3}T^2\left\{
c_A+s_F
+
{c_A^2\alpha_s\over3\pi}\left(
{5\over4}+{11\over2}\gamma_E+{11\over2}\log{\hat{\mu}\over2}
\right)
+{c_As_F\alpha_s\over\pi}\left[
{3\over4}-{4\over3}\log2
\right.\right.\\ &&\left.\left.
\!\!\!\!\!\!+{7\over6}\left( \gamma_E
+\log{{\hat{\mu}\over2}}\right)\right]
+{s_F^2\alpha_s\over\pi}\left(
{1\over3}-{4\over3}\log2-{2\over3}\gamma_E
-{2\over3}\log{{\hat{\mu}\over2}}
\right)
-{3\over2}{s_{2F}\alpha_s\over\pi}
\right\}\! .
\label{bnmass}
\eqa
This mass can be interpreted as the contribution to the Debye mass from the 
hard scale $T$ and is well defined and 
gauge invariant order-by-order in perturbation theory.
However, beyond NLO, it will also depend on factorization scale that
separates the hard scale and the soft scale $gT$.
For the quark mass, here we choose $m_q=0$.  

The final NNLO results are very insensitive to whether one chooses a
perturbative mass prescription for $m_q$ or $m_q$ = 0; 
however, convergence is improved with
the choice $m_q=0$.  A detailed presentation of the full calculation
of the NNLO thermodynamic potential and the 
dependence of our final results on the mass prescriptions for $m_D$ and $m_q$
is forthcoming in a longer paper~\cite{finalhtl}.

\section{Results}

\begin{figure}[t]
\begin{center}
\includegraphics[width=6.6cm]{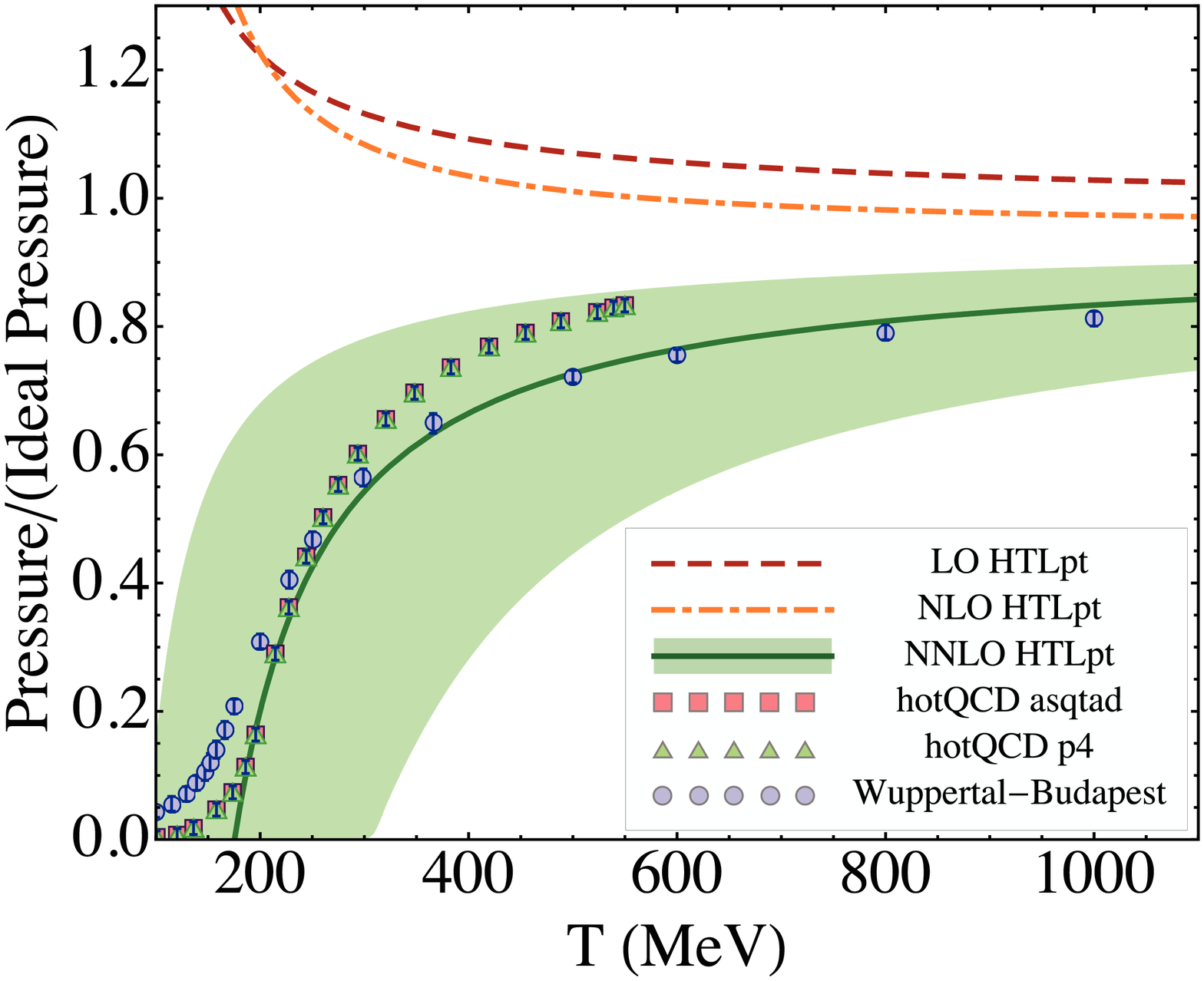}
\hspace{2mm}
\includegraphics[width=6.6cm]{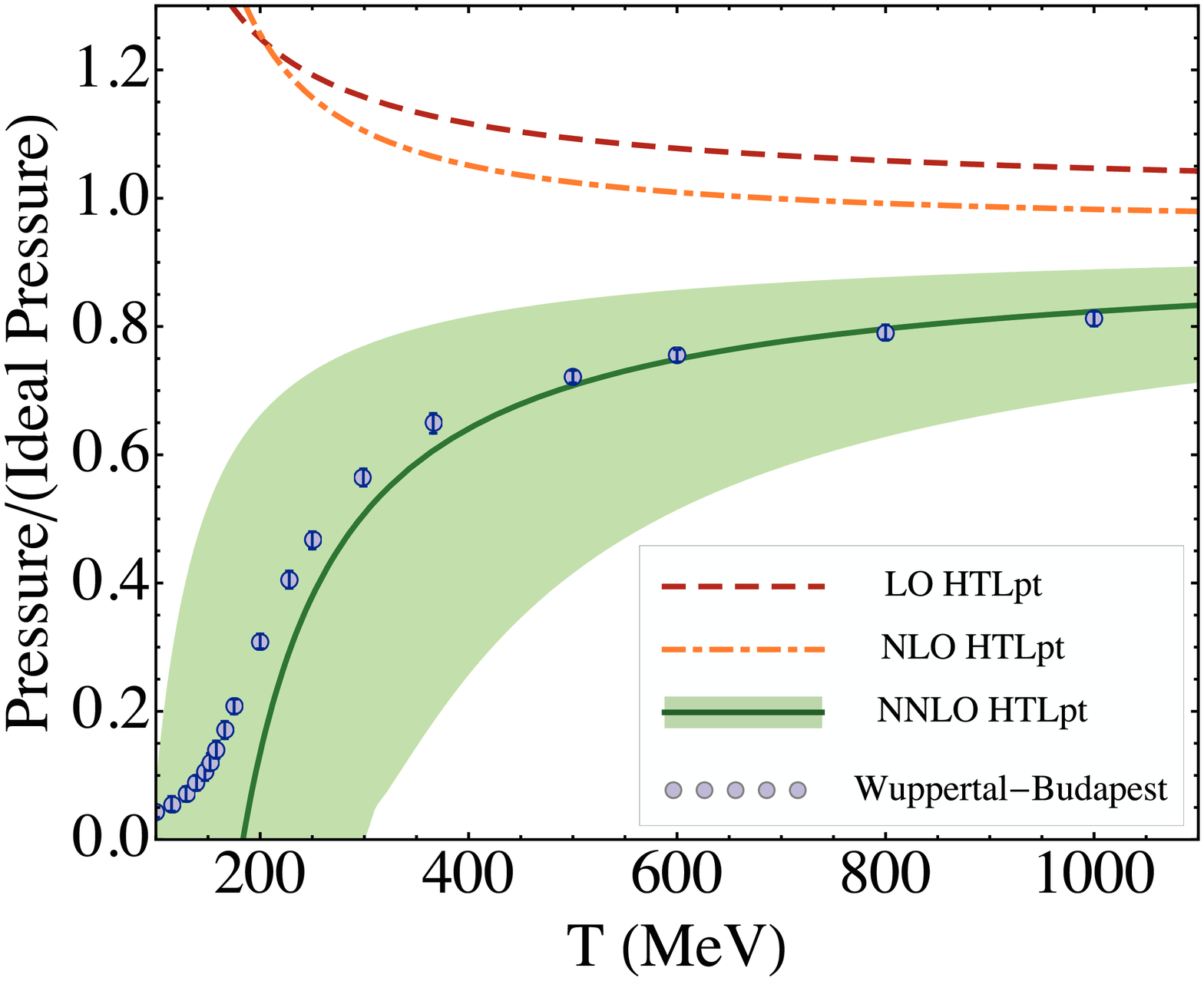}
\end{center}
\caption{Comparison of LO, NLO, and NNLO predictions for the scaled 
pressure for $N_f=2+1$ (left panel) and
$N_f=2+1+1$ (right panel) with lattice data
from Cheng et al.~\cite{hotqcd1} and 
Borsanyi et al. \cite{borsy}.
We use $N_c=3$, three-loop running for $\alpha_s$, $\mu=2\pi T$, 
and $\Lambda_{\overline{\rm MS}}=344$ MeV.
Shaded band shows the result of varying the renormalization scale $\mu$ by a 
factor of two around $\mu = 2 \pi T$ for the NNLO result. See main text for details.}  
\label{pressure}
\end{figure}

In Fig.~\ref{pressure}, we show the normalized pressure for $N_c=3$ 
and $N_f=2+1$ (left panel), and $N_c=3$ and $N_f=2+1+1$ 
(right panel) as a function
of $T$.
The results at LO, NLO, and NNLO use the BN mass given by 
Eq.~(\ref{bnmass}) as well as $m_q=0$.
For the strong coupling constant $\alpha_s$, we used three-loop 
running~\cite{partdata}
with $\Lambda_{\overline{\rm MS}}=344$ MeV which for $N_f=3$
gives $\alpha_s({\rm 5\;GeV}) = 0.2034$~\cite{McNeile:2010ji}.
The central line is evaluated with the renormalization scale $\mu = 2 \pi T$
which is the value one expects from effective field theory calculations \cite{BN,Laine:2006cp}
and the band represents a variation of $\mu$ by a factor of two around this scale.

The lattice data from the Wuppertal-Budapest collaboration
uses the stout action and have been continuum extrapolated by averaging
the trace anomaly measured using their two smallest lattice spacings
corresponding to $N_\tau = 8$ and $N_\tau = 10$
\cite{borsy}.\footnote{We note that the Wuppertal-Budapest group has published a few
data points  for the trace anomaly with $N_\tau =12$ and within statistical error bars these are consistent with
the published continuum extrapolated results.}
Using standard lattice techniques, 
the continuum-extrapolated pressure is computed from an integral of the trace anomaly.
The lattice data from the hotQCD collaboration are their $N_\tau = 8$ results
using both the asqtad and p4 actions~\cite{hotqcd1}.  The hotQCD results have not been continuum 
extrapolated and the error bars correspond to only statistical errors and do not factor in the systematic error
associated with the calculation which, for the pressure, is estimated by the hotQCD collaboration to be between 5 and 10\%.
We note that there are hotQCD results for physical light quark masses \cite{hotqcd2}; however,
these are available only for temperatures below 260 MeV and the results are 
very close to the results shown
in the figures so we do not include them here.  

As can be seen from Fig.~\ref{pressure} the successive HTLpt approximations represent an
improvement over the successive approximations coming from a naive weak-coupling expansion;
however, as in the pure-glue case~\cite{pg}, the NNLO result represents a significant correction to 
the LO and NLO results.  That being said the NNLO HTLpt result agrees quite well with the available
lattice data down to temperatures  on the order of $2\,T_c \sim 340$ MeV for both $N_f=3$ 
(Fig.~\ref{pressure} left) and $N_f=4$ (Fig.~\ref{pressure} right).  Below these temperatures
the successive approximations give large corrections with the correction from NLO to NNLO 
reaching 100\% near $T_c$.

\begin{figure}[t]
\begin{center}
\includegraphics[width=6.6cm]{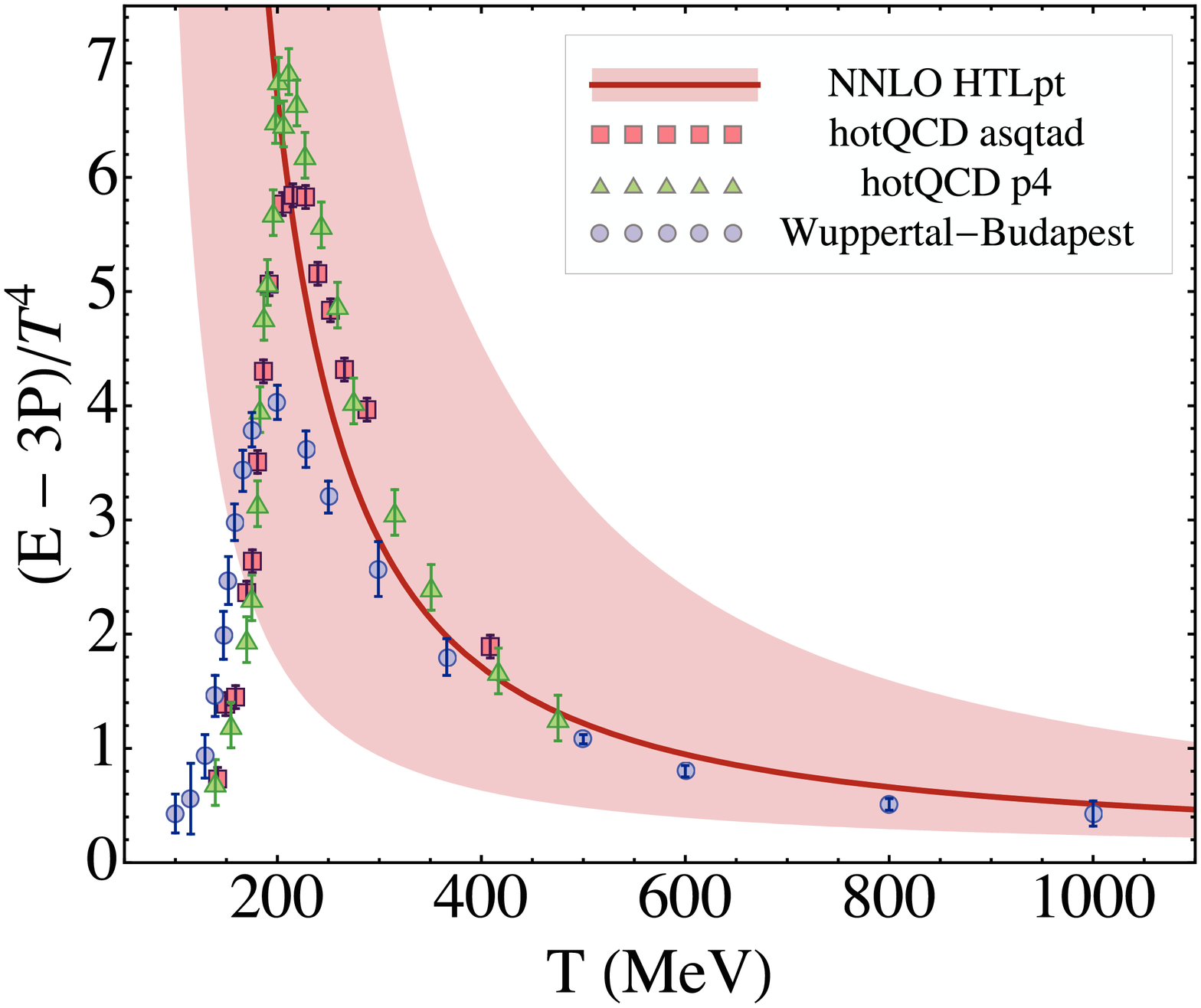}
\hspace{2mm}
\includegraphics[width=6.6cm]{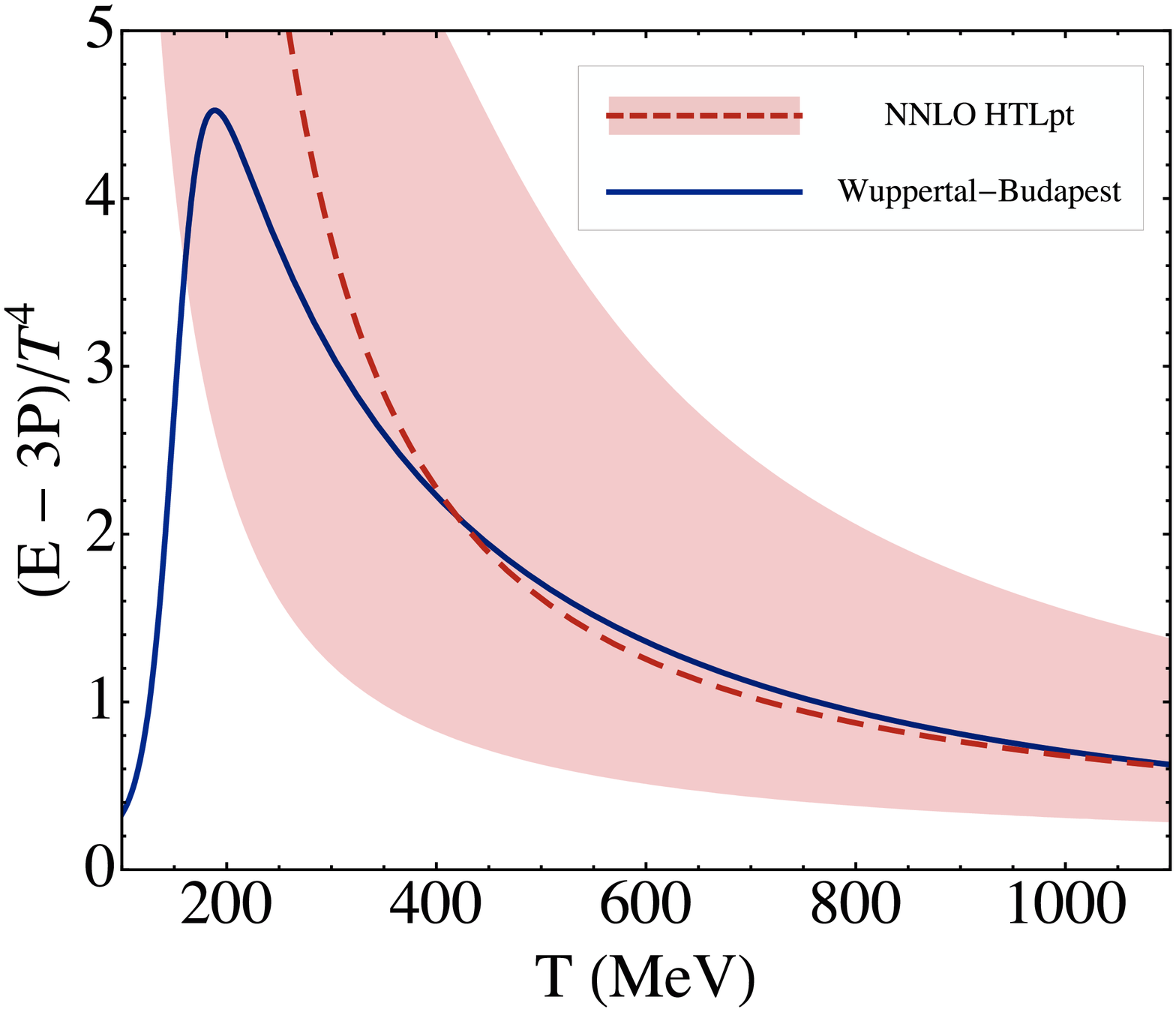}
\end{center}
\caption{Comparison of NNLO predictions for the scaled trace
anomaly with $N_f=2+1$ (left panel) and
$N_f=2+1+1$ fermions (right panel) lattice data
from Cheng et al.~\cite{hotqcd1} and 
Borsanyi et al. \cite{borsy}.
We use $N_c=3$, three-loop running for $\alpha_s$, $\mu=2\pi T$, 
and $\Lambda_{\overline{\rm MS}}=344$ MeV.   
Shaded band shows the result of varying the renormalization scale $\mu$ by a 
factor of two around $\mu = 2 \pi T$. See main text for details.
}
\label{trace2}
\end{figure}

In Fig.~\ref{trace2}, we show the NNLO approximation to the trace anomaly (interaction measure) 
normalized to $T^4$ as a function of $T$ for $N_c=3$ and $N_f=3$ (left panel) and for $N_c=3$
and $N_f=4$ (right panel).  In the left panel we show data from both the Wuppertal-Budapest
collaboration and the hotQCD collaboration taken from the same data sets displayed in
Fig.~\ref{pressure} and described previously.  In the case of the hotQCD results we note that the
results for the trace anomaly using the p4 action show large lattice size effects at all temperatures
shown and the asqtad results for the trace anomaly show large lattice size effects for $T \gsim 200$ MeV.
In the right panel we display a parameterization (solid blue curve) of the trace anomaly for $N_f=4$ published by 
the Wuppertal-Budapest collaboration \cite{borsy} since the individual data points were not published.  In both 
the left and right panels we see very good agreement with the available lattice data down 
to temperatures on the order of $T \sim 2\,T_c$.

\section{Summary and outlook}

We have reviewed recent results for the LO, NLO, and NNLO thermodynamic
functions for SU($N_c$) Yang-Mills theory with $N_f$ fermions using
HTLpt. We
compared our predictions 
with lattice data for $N_c=3$ and $N_f \in \{3,4\}$
and found that HTLpt is consistent with available lattice
data down to $T\sim2\,T_c$ for 
the pressure and the trace anomaly. This is in line with expectations since
one is expanding about the trivial vacuum $A_{\mu}=0$ and therefore neglects
the approximate 
center symmetry $Z(N_c)$. Close to the deconfinement transition, 
it is essential to
incorporate this symmetry~\cite{zenter}.

Comparing our results with the NNLO results of pure Yang-Mills~\cite{pg}, 
we find that including the quarks gives much better agreement with lattice data.
This is not unexpected since fermions are ``perturbative" in the sense that
they decouple in the dimensional-reduction step of effective field theory.

As was the case with pure Yang-Mills we found that the variational 
solution for the Debye mass $m_D$ is complex and we therefore chose 
instead to use the perturbative mass parameter from EQCD together 
with $m_q=0$. Whether the complexity of the variational Debye mass 
is due to the additional expansion in $m_D/T$ and $m_q/T$ is impossible
to decide at this stage. 
We also found that there was a large correction
going from NLO to NNLO. Unfortunately, due to the magnetic mass 
problem it is impossible to go to N$^3$LO to see whether the problem 
persists without supplementing our calculation with input from three-dimensional
lattice calculations.

In closing, we emphasize that HTLpt provides a gauge invariant 
reorganization of perturbation theory for calculating static and dynamic quantities in thermal field
theory. Given the good agreement with lattice data for thermodynamics, it would be interesting
to apply HTLpt to the calculation of  real-time quantities at temperatures that are relevant for LHC.

\vspace{-3mm}
\section*{Acknowledgments}
M. Strickland would like to thank the Yukawa Institute for Theoretical Physics, Kyoto University, 
for their hospitality and for the invitation to present this work at the ``High Energy Strong 
Interactions 2010'' symposium.  N.~Su was supported by the Frankfurt International Graduate
School for Science and Helmholtz Graduate School for Hadron and Ion Research. 
M. Strickland was supported by the Helmholtz International Center for FAIR 
LOEWE program.

\end{document}